\documentclass[sigconf]{acmart}

\usepackage{booktabs} 
\usepackage[ruled]{algorithm2e} 
\usepackage{amssymb}
\usepackage{xcolor}
\usepackage[most]{tcolorbox}
\usepackage{todonotes}
\usepackage{listings}
\usepackage{amsmath}
\usepackage{booktabs}
\usepackage{enumitem}
\usepackage{algpseudocode}
\usepackage{multirow}
\usepackage{url}
\usepackage{framed}
\usepackage{graphicx}
\usepackage{caption}
\usepackage{subcaption}
\usepackage{balance}
\usepackage{adjustbox}
\usepackage{tabularx,ragged2e,booktabs}

\setcopyright{none}
\setcopyright{acmcopyright}

\copyrightyear{2019} 
\acmYear{2019} 
\setcopyright{acmcopyright}
\acmConference[SIGIR '19]{Proceedings of the 42nd International ACM SIGIR Conference on Research and Development in Information Retrieval}{July 21--25, 2019}{Paris, France}
\acmBooktitle{Proceedings of the 42nd International ACM SIGIR Conference on Research and Development in Information Retrieval (SIGIR '19), July 21--25, 2019, Paris, France}
\acmPrice{15.00}
\acmDOI{10.1145/3331184.3331353}
\acmISBN{978-1-4503-6172-9/19/07}

\begin{document}

\title{Revealing the Role of User Moods in Struggling Search Tasks}
\author{Luyan Xu}
\affiliation{%
  \institution{DEKE Lab, \\ Renmin University of China}
  \city{Beijing}
  \state{China}
  }
\email{xuluyan@ruc.edu.cn}

\author{Xuan Zhou}
\affiliation{%
  \institution{School of Data Science \& Engineering, \\ East China Normal University}
  \city{Shanghai}
  \state{China}
  }
\email{zhou.xuan@outlook.com}

\author{Ujwal Gadiraju}
\affiliation{%
  \institution{L3S Research Center, \\ Leibniz Universit\"at Hannover}
 \city{Hannover}
  \state{Germany}
}
\email{gadiraju@L3S.de}


\begin{abstract}
User-centered approaches have been extensively studied and used in the area of struggling search. Related research has targeted key aspects of users such as user satisfaction or frustration, and search success or failure, using a variety of experimental methods including laboratory user studies, in-situ explicit feedback from searchers and by using crowdsourcing. Such studies are valuable in advancing the understanding of search difficulty from a user's perspective, and yield insights that can directly improve search systems and their evaluation. However, little is known about how user moods influence their interactions with a search system or their perception of struggling. In this work, we show that a user's own mood can systematically bias 
the user's perception, and experience while interacting with a search system and trying to satisfy an information need.  People who are in \textit{activated-(un)pleasant} moods tend to issue more queries than people in \textit{deactivated} or \textit{neutral} moods. Those in an unpleasant mood perceive a higher level of difficulty. Our insights extend the current understanding of struggling search tasks and have important implications on the design and evaluation of search systems supporting such tasks. 
\end{abstract}

%
%



\maketitle

\section{Introduction}
\label{sec:intro}
Methods have recently been developed to understand users' struggle during search experiences and help them cope with the entailing search difficulty. As a subject of scholarly attention, user-centered approaches have been studied in many different contexts of struggling search, with a variety of people and exploring a broad array of insights in user-system interaction, user behavior analysis, and task difficulty evaluation. 
However, few studies have taken into account the role that a user's mood plays in these processes.

Previous works have not explored the impact of user moods on their perception of search struggle or satisfaction.
Recently, researchers found the `difficulty' in struggling search to be a function of a searcher's effort and gain, based largely on subjective feelings~\cite{wildemuth2014untangling}. We draw inspiration from this work to analyze the role of user moods in search. 

To investigate whether and how user moods affect their search performance and perception of difficulty in struggling search, we conducted a crowdsourced user study. 
We recruited participants (\textit{N=284}) from a popular crowdsourcing platform called Figure-Eight\footnote{\url{http://www.figure-eight.com/}}. We used \textit{Pick-A-Mood} (PAM), a character-based pictorial scale, to gather self-reported user moods from workers~\cite{desmet2016mood}. We investigated the effects of users' moods on their perception of task difficulty, and analyzed their search behavior. We compared user performance in typical information retrieval (IR) tasks and that in struggling text retrieval tasks (SST).  In this paper, we address the following research questions:

\textbf{RQ1:} What moods are participants in when they begin their search session?  
We gathered user moods using PAM, and analyzed these across 4 distinct mood categories. 

\textbf{RQ2:} What is the effect of mood on participants' search behavior during the struggling search process? We logged and analyzed user search behavior including queries, clicks and dwelling time. 

\textbf{RQ3:} What is the effect of mood on participants' perception of task difficulty during the struggling search process? Through a quantitative analysis of the difficulty scores which users assigned to tasks, we investigated the influence of mood on users' perception of task difficulty.

\textbf{RQ4:} What is the effect of mood on typical IR tasks in comparison to struggling search tasks? We also measured the impact of mood on typical IR tasks, drawing comparisons to our findings with respect to struggling search tasks.


\section{Related Literature}


Although both \textit{mood} and \textit{emotion} are valenced affective responses, prior work has elaborately discussed the difference between the two \cite{desmet2016mood}. Firstly, moods last longer than emotions \cite{beedie2005distinctions,verduyn2011relation}. Secondly, emotions are always targeted towards an event, person or object, while moods are globally diffused \cite{frijda1994varieties}. Emotions are triggered by explicit causes and monitor our environment, while moods have combined causes and monitor our internal state \cite{morris2012mood}. 
Further, emotions are elicited by threats or opportunities \cite{frijda1994varieties}, while moods are responses to one's overall position in general \cite{prinz2004gut}. However, note that moods and emotions are not entirely independent; they interact with each other dynamically. Accumulated emotions can lead to specific moods, and moods can lower the degree of emotional arousal \cite{davidson1994emotion}.

Crowdsouring is increasingly being used to build reliable hybrid human-machine information systems \cite{demartini2017introduction} and to run human-centered experiments \cite{YGHRKD_sigir2018}. It has emerged as a feasible approach to gather reliable relevance labels in the context of IR evaluation. Platforms, such as FigureEight~\footnote{Figure8 -- \url{http://figure-eight.com/}} or Amazon's Mechanical Turk (AMT)~\footnote{\url{http://www.mturk.com/}}, enable the gathering of vast amount of data from a large population of workers within a short period of time and at a relatively low cost \cite{gadiraju2017crowdsourcing}.
In this paper, we build on these substantial prior works \cite{verduyn2011relation,frijda1994varieties,morris2012mood,prinz2004gut,davidson1994emotion,beedie2005distinctions}, 
that have established an understanding of \textit{moods} and \textit{emotions} to unearth 
analyze the role they play on workers of microtask crowdsourcing platforms.

\section{Method and Setup}
\label{sec:method}

\subsection{Measuring User {Moods}}

\begin{figure}[h]
    \vspace{-.5em}
   \centering
   \includegraphics[width=0.35\textwidth]{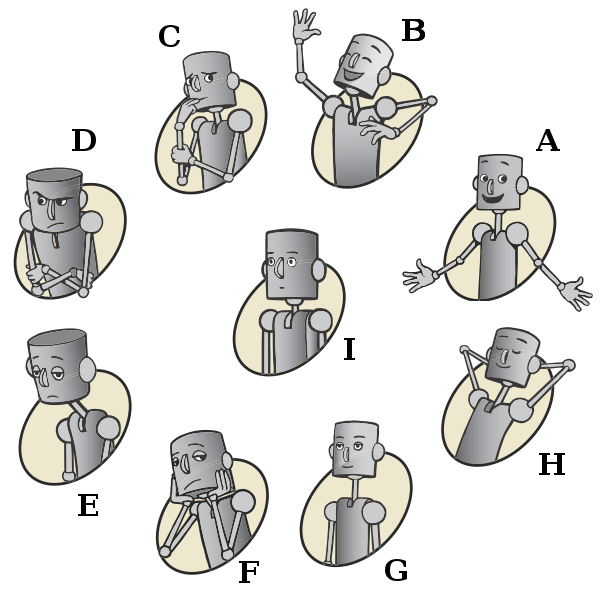}
   \vspace{-.5em}
   \caption{\textit{Pick-A-Mood} scale to measure the self-reported mood of users before they enter the \textit{TaskGenie} framework.}
   \label{fig:pam}
\end{figure}

To measure the \textit{mood} of crowd workers in an intuitive and easy manner, we use \textit{Pick-A-Mood} (PAM), a character-based pictorial scale for reporting moods \cite{desmet2016mood}. Compared to other measures, this is ideal for the microtask crowdsourcing context where time is of essence, since it was specifically made to be suitable for design research applications in which people have little time or motivation to report their moods. The scale has been tested with a general population (people from 31 different nationalities in the validation study), revealing that the expressions presented by the visual characters are correctly interpreted (see Figure \ref{fig:pam}). {PAM has been used in a variety of research settings including quality of experience research \cite{villa2013investigating} and in education \cite{vandevelde2016design}, illustrating the robustness of the tool.}

\subsection{Tasks}
To analyze how mood effects users' search behavior in struggling search tasks (SSTs), we formulated 10 struggling text retrieval tasks from Wikipedia using a method from previous study~\cite{xu2019generating}. We made sure that the first interaction of the user with the system would not directly yield the answer to the information need~\cite{kules2008creating}. The generated tasks are characterized by an open task description, accompanied by uncertainty and ambiguity as postulated by Kules and Capra~\cite{kules2009designing}. In addition, the tasks were designed such that searchers had to search for multiple aspects describing the information need~\cite{kules2008creating}. To observe the difference between typical IR tasks and SST in terms of the effect of user mood, we picked another 10 typical IR tasks from previous work by Gadiraju et al. that was based on the \textit{TREC 2014 Web Track} dataset\footnote{\url{http://www.trec.nist.gov/act\_part/tracks/web/web2014.topics.txt}} \cite{gadiraju2018analyzing}. Table \ref{tasks} presents samples of the selected typical IR tasks and the generated struggling text retrieval tasks. All tasks are made publicly available\footnote{https://bitbucket.org/ielool/mood\_taskset/}. 

\begin{table}[h]
\vspace{-.5em}
\caption{\label{tasks} Examples of Traditional IR tasks and SST\vspace{-.75em}}
\resizebox{.48\textwidth}{12.5mm}{
\begin{tabular}{clll}
\toprule
\textbf{TASK\_TYPE}                                                         & \multicolumn{3}{c}{\textbf{Sample of Tasks Generated in the Lab}}                                                                                                                                                                                            \\ \midrule
\multirow{2}{*}{\begin{tabular}[c]{@{}c@{}}Typical\\ IR Tasks\end{tabular}} & \multicolumn{3}{l}{Which astronomer is the Hubble Space Telescope named after?}                                                                                                                                                      \\
                                                                            & \multicolumn{3}{l}{Which is the highest summit of the Rocky Mountains?}                                                                                                                                                              \\
\multicolumn{1}{l}{}                                                        & \multicolumn{3}{l}{}                                                                                                                                                                                                                 \\
\multirow{2}{*}{SST}                                                      & \multicolumn{3}{l}{\begin{tabular}[c]{@{}l@{}}Did the fall of Dien Bien Phu upset the balance of forces present \\ in Indochina in 1954?\end{tabular}}                                                                               \\
                                                                            & \multicolumn{3}{l}{\begin{tabular}[c]{@{}l@{}}Which bonds do nucleases hydrolyze to cut DNA strands?\end{tabular}} \\ \bottomrule
\end{tabular}}
\vspace{-1em}
\end{table}

\subsection{Experimental Setup and Data Collection}

We built \textit{TaskGenie}\footnote{http://waps.io/study/?uid=123}, a customized online search engine capable of logging user behavior, on top of the Wikipedia Search API\footnote{https://en.wikipedia.org/w/api.php}. By default, to ensure that retrieved documents are consistent in their credibility and coverage, we specify Wikipedia as the target search domain. We logged user activity on the platform including queries, clicks, users' dwelling time and their perception of task difficulty using PHP/Javascript and the jQuery library.

\begin{figure}[h]
    \vspace{-1.5em}
   \centering
   \includegraphics[width=0.45\textwidth]{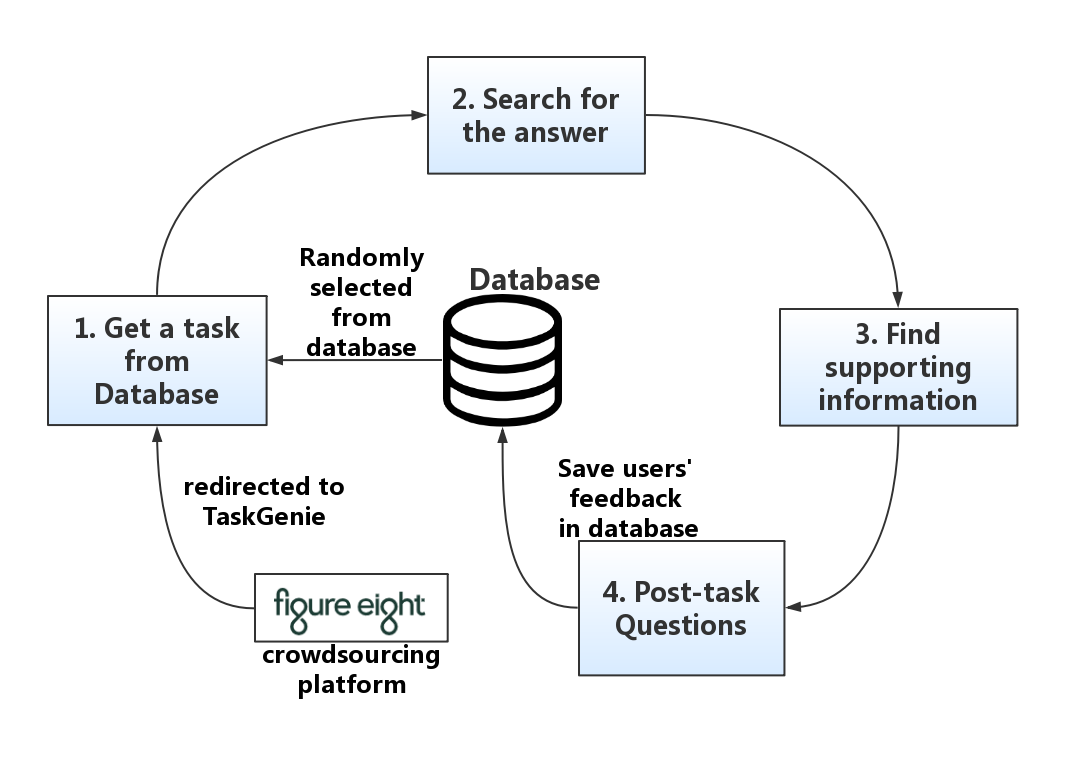}
   \vspace{-1em}
   \caption{ Workflow of participants in the experimental setup orchestrating informational search sessions within the \textit{TaskGenie} framework.}
   \label{fig:para}
   \vspace{-1em}
\end{figure}

We recruited 300 participants (94 female, 206 male, with their age ranging from 18 to 57) from FigureEight~\footnote{{Figure8 -- \url{http://figure-eight.com/}}}, a premier crowdsourcing platform. We restricted the participation to workers from English-speaking countries to ensure that they understood the task and instructions adequately. To ensure reliability of the resulting data, we restricted the participation to \textit{Level-3 workers~\footnote{\textit{Level-3 contributors} on Figure8 
are workers of the highest quality.}}.
At the onset, workers were informed that the task entailed `searching the web to find information'. Workers willing to participate were first asked to respond to a few general questions pertaining to their gender, age and select a mood that could describe their state, before being redirected to the aforementioned external platform, \textit{TaskGenie}. 
Workers were randomly assigned either a 
a struggling search task or a typical IR task. 
 During the task completion process, all interactions of the users within the \textit{TaskGenie} framework were logged. After finishing the task, searchers were asked to give feedback from the following perspectives - (1) \textit{Task Comparison} (whether or not the users found the question difficult in comparison to their usual experience); (2) \textit{Task Difficulty Score} (how difficult / complex the users found the question to be on a sliding scale of 1 to 100). We divide the task difficulty scale into five equal parts using the following labels - \textit{Easy} (1-20), \textit{Moderate} (21-40), \textit{Challenging} (41-60), \textit{Demanding} (61-80), \textit{Strenuous} (81-100).
Figure~\ref{fig:para} depicts the workflow of participants in the experimental setup orchestrating informational search sessions.

To ensure the reliability of responses and the behavioral data thus produced in the search sessions, we filtered out 16 workers who entered no queries or didn't finish the whole process. We got 284 workers of which 174 solved a struggling search tasks and 110 solved typical IR tasks. The analysis and results presented hereafter are based on these 284 workers.

\section{Results and Analysis}
\label{sec:results_analysis}
In this section, we aim to analyze the effect of moods on users' searching process in SSTs and typical IR tasks based on a variety of aspects. Note that to control for Type-I error inflation in our multiple comparisons (presented in this section), we used the Holm Bonferroni correction for family-wise error rate (FWER) \cite{holm1979simple}, at the significance level of $\alpha$ $<$ .05.

\subsection{What Mood Were the Users In?}
The 8 non-neutral moods measured by PAM, can be grouped into four mood categories \cite{watson1985toward}; \textbf{activated-pleasant} (\textit{excited, cheerful}), \textbf{deactivated-pleasant} (\textit{relaxed, calm}), \textbf{activated-unpleasant} (\textit{tense, irritated}), and \textbf{deactivated-unpleasant} (\textit{bored, sad}). 

Among the 284 workers, 147 workers (51\%) were found to be in \textit{activated-pleasant} moods (95 were cheerful and 52 were excited), 62 workers (22\%) were in \textit{deactivated-pleasant} moods (35 were calm, 27 were relaxed). 20 workers (7\%) were in \textit{activated-unpleasant} moods (16 were tense, 4 were irritated) and 33 (12\%) were in \textit{deactivated-unpleasant} moods (19 were sad, 14 were bored). 22 workers (8\%) claimed to be in a neutral mood. Thus, more than half of the workers were in activated-pleasant moods, while there were nearly 30\% of the workers who were in neutral or unpleasant moods.

\subsection{How Mood Affects Struggling Search?}

We analyzed how different kinds of mood influenced the search behavior of users (i.e. query, clicks, duration, user perception of difficulty) in the informational search sessions corresponding to the 10 struggling search tasks (SST). 
\newline
\textbf{Queries.} During the search sessions corresponding to the 20 tasks, users issued over two queries on average. To understand the effect of users moods on the number of queries issued by the users, we conducted a one-way ANOVA. We found that people who were in activated moods  issued more queries than people in deactivated or neutral moods. Results revealed a significant effect of moods on the number of queries issued by users at the $p < .001$ level, $F(8, 165) = 4.38$. Post-hoc comparisons using the Tukey-HSD test revealed that the number of issued queries in activated-pleasant mood (`\textit{cheerful}' and `\textit{excited}') were found to be significantly more than those pertaining to the mood of `\textit{neutral}' at the $p<.001$ level. 
\newline
\textbf{Clicks.} We analyzed the clicks of users on results corresponding to each of the queries they entered within search sessions. We note that users clicked on just over 355 links on the search results and on more than 1 result link per query on average. We conducted a one-way ANOVA to investigate the effect of different moods on the number of clicks fired by users. We did not find a significant difference in the number of clicks fired by users across the 9 types of moods. We also found no significant linear relationship between the mood and the number of clicks using Pearson's \textbf{R}.
\newline
\textbf{Duration.} We analyzed the session length\footnote{For a given topic and user, we measured the session length as the time from which the first query was entered in \textit{TaskGenie} by the user after the calibration test, until the time at which the last web page accessed by the user was active before the post-session test. Note that users were allowed to carry out only one search session.} of users in informational search sessions corresponding to the different moods. We found that the average session length of solving a task is 318s (\textit{M = 318, SD = 147}) long. To understand the effect of the 9 types of moods on the session length exhibited by users, we conducted a one-way ANOVA. Results revealed no significant effect of the moods on sessions length. We also analyzed the difference between session length of users and the positive / negative moods. We found no significant difference between these variables, suggesting that types of mood do not directly influence the the session length of users.
\newline
\textbf{Perception of Difficulty.} We analyzed users' perception of their search based on the difficulty ratings we collected at the end of their search session on a sliding scale of 1-100 as described earlier. After being assigned a task at random some users chose to change their tasks once. We found that users in activated-pleasant (\textit{excited} and \textit{cheerful}) moods were more inclined to switch the tasks they were assigned. Results of a two-tailed T-test confirmed this significant difference: $t(173) = 2.81$, at $p < .01$ level. From this we note that searchers in activated-pleasant moods tend to be more active when interacting with the search engine. 72\% of the users who completed the tasks claimed that the tasks they completed were difficult tasks in comparison to their typical search experience. To understand whether users moods influenced their perception in this self-reported comparison, we conducted a one-way between users ANOVA. We found no significant effect of mood on how participants compared the struggling search tasks and their typical searching experience. 
On average, the perceived task difficulty corresponding to the 20 tasks was found to be 59.7; \textit{challenging} (\textit{M = 59.7, SD = 21.9}). To understand the effect of the 9 different moods on the perception of difficulty, we conducted a one-way between users ANOVA. Results revealed a significant effect of user moods on the \textit{Task Difficulty Score} across the 9 types of moods at $p < .01$ level; $F(8, 165) = 2.38$. Post-hoc comparisons using the Tukey-HSD test revealed that people in \textit{activated-unpleasant} and \textit{deactivated-unpleasant} mood reported a  higher perceived difficulty score. In addition, the difficulty score pertaining to users in a `sad' (deactivated-pleasant) mood  was found to be significantly higher that that of those in a `cheerful' (activated-pleasant) mood, at $p < 0.01$ level.

\subsection{Impact of Moods on IR tasks vs. SST}
We also investigated the effect of mood on users carrying out typical IR tasks, to draw a comparison with those carrying out SST. 

We found that on average users collectively fired one to two queries across each of the 10 typical IR tasks. We conducted a one-way ANOVA to investigate the effect of moods on the number of queries fired in the typical IR tasks. Results showed a significant effect of different kinds of moods on the number of queries issued by users at $p < .01$ level; $F(8, 101) = 3.42$. Post-hoc comparisons using the Tukey-HSD test revealed that the number of queries issued by workers in an `\textit{irritated}' mood was significantly more than those in a `\textit{neutral}' mood at $p < .05$ level. Similar to how moods effect SST, in typical IR tasks we found that people who were in \textit{activated-pleasant} and \textit{activated-unpleasant} moods tend to issue more queries than people in a neutral mood.  Participants' perception of task difficulty for typical IR tasks was found to be 35 (\textit{M = 35, SD = 25}).
In contrast to what we found in struggling search tasks, where user moods affected their perception of difficulty of the SSTs, we found that user moods do not affect the perception of difficulty in typical IR tasks. This suggests that if 
the information need can be relatively easily satisfied, user moods do not affect their perception of task difficulty.

We note that users in total fired 141 clicks and chained around 1 link on average, and spent around 220s on solving one typical IR task. We found no significant effect of mood on the number of clicks and duration time through one-way ANOVA test. We also found no significant linear relationship between the mood and these two parameters using Pearson's \textbf{R}. 

\subsection{Discussion} 

We analyzed the user behavioral data to address the four RQs. 
In our study, we found that more than half of the participants' were in activated-pleasant moods, although users were spread across the 9 moods in general. We found that user moods can affect the number of queries issued. Users who were in activated-(un)pleasant moods issued more queries than those in deactivated or neutral moods (i.e. users in activated mood either intentionally explored the results or struggled to find the information they were seeking). In struggling search tasks, we found that mood can affect users' perception of the task difficulty: people in activated-pleasant moods were more active in tackling a task; people in negative moods perceived a higher difficulty and assigned higher task difficulty scores. This suggests that users moods do play a role in the `struggle' they experience while searching in SSTs. Considering user moods would be helpful in eliminating mood related biases in both the experimental setup and analysis of struggling search studies. Furthermore, the comparison of the effects of mood between IR tasks and struggling search tasks showed that mood has a greater overall impact on users' perception of task difficulty in case of struggling search tasks than in typical IR tasks. This presents us a new perspective from which we can distinguish struggling search tasks and typical IR search tasks through user logs analysis.

\section{Conclusion}
The primary goal of our research was to explore \textit{whether and how} users' mood influences their search behavior. Based on the \textit{Pick-A-Mood} pictorial scale, we investigated the effect of mood on participants' search behavior (i.e. queries, clicks and dwelling time), and on their perception of task difficulty. We also compared the effect of mood on typical IR tasks and SSTs. Our findings showed that in struggling search tasks user moods can affect the number of queries fired by searchers and their perception of task difficulty. We expect that our findings will help in reducing mood-related bias in the evaluation, analysis and study of user behavior in SSTs. 
\balance

\bibliographystyle{ACM-Reference-Format}
\bibliography{references}

\end{document}